\documentclass[12pt,a4paper,english,amssymb,nofootinbib,superscriptaddress]{revtex4}
\usepackage{lmodern}
\usepackage{lmodern}
\usepackage[T1]{fontenc}
\usepackage[latin9]{inputenc}
\usepackage{color}
\usepackage{babel}
\usepackage{amsmath}
\usepackage{amssymb}
\usepackage{esint}
\PassOptionsToPackage{normalem}{ulem}
\usepackage{ulem}
\usepackage{subscript}
\usepackage[unicode=true,pdfusetitle,
 bookmarks=true,bookmarksnumbered=false,bookmarksopen=false,
 breaklinks=false,pdfborder={0 0 1},backref=false,colorlinks=true]
 {hyperref}
\hypersetup{
 linkcolor=blue,citecolor=blue}

\makeatletter

\pdfpageheight\paperheight
\pdfpagewidth\paperwidth

\@ifundefined{textcolor}{}
{%
 \definecolor{BLACK}{gray}{0}
 \definecolor{WHITE}{gray}{1}
 \definecolor{RED}{rgb}{1,0,0}
 \definecolor{GREEN}{rgb}{0,1,0}
 \definecolor{BLUE}{rgb}{0,0,1}
 \definecolor{CYAN}{cmyk}{1,0,0,0}
 \definecolor{MAGENTA}{cmyk}{0,1,0,0}
 \definecolor{YELLOW}{cmyk}{0,0,1,0}
}

\usepackage{babel}
\usepackage{babel}
\@ifundefined{definecolor}{color}{}
\usepackage{babel}

\usepackage{latexsym}\usepackage{bm}

\makeatother

\begin{document}

\title{Gravity Waves in Three Dimensions}

\author{Metin Gürses}

\email{gurses@fen.bilkent.edu.tr}

\affiliation{{\small{}Department of Mathematics, Faculty of Sciences}\\
 {\small{}Bilkent University, 06800 Ankara, Turkey}}

\author{Tahsin Ça\u{g}r\i{} \c{S}i\c{s}man}

\email{tahsin.c.sisman@gmail.com}

\affiliation{Department of Astronautical Engineering,\\
 University of Turkish Aeronautical Association, 06790 Ankara, Turkey}

\author{Bayram Tekin}

\email{btekin@metu.edu.tr}

\affiliation{Department of Physics,\\
 Middle East Technical University, 06800 Ankara, Turkey}

\date{\today}
\begin{abstract}
We find the explicit forms of the anti-de Sitter plane, anti-de Sitter
spherical, and pp waves that solve both the linearized and exact field
equations of the most general higher derivative gravity theory in
three dimensions. As a sub-class, we work out the six derivative theory
and the critical version of it where the masses of the two spin-2
excitations vanish and the spin-0 excitations decouple.

\tableofcontents{} 
\end{abstract}
\maketitle

\section{Introduction\label{sec:Introduction}}

Due to the nonlinearity of Einstein's equations, it is highly difficult
to find exact solutions. This is even more so in modified gravity
theories where more powers of curvature added to the Einstein-Hilbert
action to make the theory better behaved in the UV region. Therefore,
it is quite important to find exact solutions of higher derivative
gravity theories. Especially, for the purposes of the anti-de Sitter/conformal
field theory correspondence, it is highly desirable to find some ``neighboring''
solutions to the AdS spacetime. This work started with the purpose
of providing some AdS-related solutions to generic gravity theories
in three dimensions. For a specific quadratic curvature gravity, called
the new massive gravity (NMG) \cite{BHT}, these kind of solutions
were studied in \cite{Ayon-Beato-Bending,Gurses-Killing,Ahmedov-Aliev-PRL,Ahmedov-Aliev-PLB}.
Recently \cite{Gurses-PRL,gur-sis-tek}, using general arguments,
we have shown that the AdS-wave and the pp-wave metrics solve the
most general gravity theory with the action in the $n$-dimensional
spacetime given as 
\begin{equation}
I=\int d^{n}x\sqrt{-g}\, F\left(g^{\alpha\beta},R_{\phantom{\mu}\nu\gamma\sigma}^{\mu},\nabla_{\rho}R_{\phantom{\mu}\nu\gamma\sigma}^{\mu},\dots,\left(\nabla_{\rho_{1}}\nabla_{\rho_{2}}\dots\nabla_{\rho_{M}}\right)R_{\phantom{\mu}\nu\gamma\sigma}^{\mu},\dots\right),\label{eq:Functional_general_theory}
\end{equation}
where $F$ is a differentiable function of its arguments. In this
work, as an explicit example to our formalism, we shall provide the
solutions of the most general sixth order theory in three dimensions.
These wave solutions can be written in the Kerr-Schild form

\begin{equation}
g_{\mu\nu}=\bar{g}_{\mu\nu}+2V\lambda_{\mu}\lambda_{\nu},\label{ks}
\end{equation}
with $\bar{g}_{\mu\nu}$ as the ``background metric'' which is the
flat Minkowski metric for the pp-waves and the AdS spacetime for the
AdS-wave metrics. The properties of the $\lambda^{\mu}$-vector are
crucial: it is a null and a geodesic vector. Namely, it satisfies
the following expressions for both $g_{\mu\nu}$ and $\bar{g}_{\mu\nu}$:
\begin{equation}
\lambda_{\mu}\lambda^{\mu}=g_{\mu\nu}\lambda^{\mu}\lambda^{\nu}=\bar{g}_{\mu\nu}\lambda^{\mu}\lambda^{\nu}=0,\label{eq:null}
\end{equation}
\begin{equation}
\lambda^{\mu}\nabla_{\mu}\lambda_{\rho}=\lambda^{\mu}\bar{\nabla}_{\mu}\lambda_{\rho}=0,\label{eq:geodesic}
\end{equation}
\begin{equation}
\nabla_{\mu}\lambda_{\nu}=\bar{\nabla}_{\mu}\lambda_{\nu}=\lambda_{(\mu}\xi_{\nu)}=\frac{1}{2}\left(\lambda_{\mu}\xi_{\nu}+\xi_{\mu}\lambda_{\nu}\right),\qquad\xi^{\mu}\lambda_{\mu}=0,\label{eq:Kundt}
\end{equation}
where $\bar{\nabla}_{\mu}$ is the covariant derivative with respect
to the background metric. The last property restricts the Kerr-Schild
metric to the Kundt class where the $\lambda^{\mu}$ vector is nonexpanding,
shear-free, and nontwisting. Due to this property, we denote this
class of metrics as Kerr-Schild-Kundt (KSK) metrics. The new vector
$\xi^{\mu}$ that appears in (\ref{eq:Kundt}) is defined via that
equation. The metric function $V$ satisfies $\lambda^{\mu}\partial_{\mu}V=0$.
Let us suppose that the most general theory is a $2N+2$ derivative
theory; namely, the highest partial derivative of the metric in the
field equations is $2N+2$. For example, Einstein's gravity has $N=0$,
any of the form $f\left({\rm Riemann}\right)$ with no derivatives
of the Riemann tensor but only quadratic and more contractions, has
$N=1$. Explicit AdS-wave solutions of these theories have been considered
before \cite{Hassaine,Gurses-Killing,Ahmedov-Aliev-PRL,Ahmedov-Aliev-PLB,Alishah,Gullu-Gurses}.
Explicit solutions of the most general $N=2$ theory, namely the six
derivative theory, have not been considered before. Here, we shall
remedy this in three dimensions for the most general theory.

We have shown that for the metrics of the form (\ref{ks}) having
the properties (\ref{eq:null}-\ref{eq:Kundt}), all curvature scalars
are constant and the scalar curvature is $R=-\frac{6}{\ell^{2}}$,
and the traceless part of the Ricci tensor, that is $S_{\mu\nu}\equiv R_{\mu\nu}-\frac{1}{3}g_{\mu\nu}R$,
reduces to the following simple expression \cite{gur-sis-tek} 
\begin{equation}
S_{\mu\nu}=-\left(\bar{\square}+\frac{2}{\ell^{2}}\right)\lambda_{\mu}\lambda_{\nu}V\equiv\lambda_{\mu}\lambda_{\nu}\mathcal{O}V,\label{tracefree}
\end{equation}
where $\ell$ is the AdS radius, $\bar{\square}$ is the Laplace-Beltrami
operator of the background metric, and the operator $\mathcal{O}$
can be found in three dimensions as
\begin{equation}
\mathcal{O}=-\left(\bar{\square}+2\xi^{\mu}\partial_{\mu}+\frac{1}{2}\xi^{\mu}\xi_{\mu}-\frac{2}{\ell^{2}}\right).
\end{equation}
The field equations of the most general $\left(2N+2\right)$-derivative
gravity theory splits into two parts: one is the trace part that determines
the AdS radius in terms of the parameters given in the action such
as the bare cosmological constant and the coefficients of the curvature
terms. The other equation is the traceless part which reads as \cite{gur-sis-tek}
\begin{equation}
\sum_{n=0}^{N}\, a_{n}\,\bar{\square}^{n}\, S_{\mu\nu}=0\label{denk0}
\end{equation}
where $a_{n}$'s ($n=0,1,2,\cdots$) are constants which are again
functions of the parameters of the theory whose proof for the AdS-spherical
wave will be given in \cite{KSK_proof}. Equation (\ref{denk0}) can
factored as

\begin{equation}
\prod_{n=1}^{N}\,\left(\bar{\square}+c_{n}\right)S_{\mu\nu}=0,\label{denk11}
\end{equation}
where $c_{n}$'s are the roots of the polynomial 
\begin{equation}
a_{N}y^{N}+a_{N-1}y^{N-1}+\cdots+a_{1}y+a_{0}=0,
\end{equation}
which are a priori complex in general. But, in order for the theory
to be free of tachyons, all the roots must be real since they are
related to the masses of the spin-2 excitations about the AdS background
through the relation

\begin{equation}
c_{n}=\frac{2}{\ell^{2}}-m_{n}^{2},~~n=1,2,\cdots,N.\label{eq:c_n_mass_relation}
\end{equation}
This can be understood as follows: once a perturbation about the AdS
background $h_{\mu\nu}\equiv g_{\mu\nu}-\bar{g}_{\mu\nu}$ is defined
as $h_{\mu\nu}\equiv2V\lambda_{\mu}\lambda_{\nu}$, with all the properties
of $\lambda_{\mu}$ and $V$ intact as in the exact solution, then
the exact solution and the perturbative solution for this particular
transverse-traceless $h_{\mu\nu}$ representing spin-2 modes become
equal. Note that the spin-0 modes cannot be obtained this way as the
full spacetime is a constant curvature spacetime, namely linearized
part of the scalar curvature is zero. For the case of the pp-wave
metrics, one takes the limit $\ell\rightarrow\infty$. Note that the
naïve counting of the degrees of freedom in terms of the metric \emph{alone}
in these higher derivative theories would take one astray: for example,
in four dimensions, one would conclude that a symmetric two-tensor,
$h_{\mu\nu}$, could have at most 10 propagating degrees of freedom.
This could only be true in a second derivative theory without any
symmetries. On the other hand, in higher derivative theories, $\partial^{p}h_{\mu\nu}$
type objects should be considered as independent fields as was done
by Pais and Uhlenbeck \cite{Pais-Uhlenbeck}. Since $S_{\mu\nu}$
satisfies (\ref{tracefree}) then (\ref{denk11}) reduces to

\begin{equation}
\prod_{n=1}^{N}\,\left(\mathcal{O}+m_{n}^{2}\right)\,\mathcal{O}V=0,\label{denk2-1}
\end{equation}
where we also used the relations 
\begin{equation}
\square\left(\phi\lambda_{\alpha}\lambda_{\beta}\right)=\bar{\square}\left(\phi\lambda_{\alpha}\lambda_{\beta}\right)=-\lambda_{\alpha}\lambda_{\beta}\left(\mathcal{O}+\frac{2}{\ell^{2}}\right)\phi,
\end{equation}
which are valid for any function $\phi$ satisfying $\lambda^{\mu}\nabla_{\mu}\phi=0$.
This also leads to $\lambda^{\mu}\nabla_{\mu}\mathcal{O}\phi=0$.
Provided that that all $m_{n}^{2}$'s are different, the most general
solution to (\ref{denk2-1}) can be written as 
\begin{equation}
V=V_{E}+\sum_{n=1}^{N}\, V_{n},\label{eq:Gen_soln}
\end{equation}
where $V_{E}$ represents the solution to the cosmological Einstein's
theory satisfying 
\begin{equation}
\mathcal{O}V_{E}=0.\label{denk3}
\end{equation}
In three dimensions, the solutions of this equation can be ``gauged
away'': namely, the metric $\bar{g}_{\mu\nu}+2V_{E}\lambda_{\mu}\lambda_{\nu}$
is that of ${\rm AdS}_{3}$. This is related to the fact that cosmological
Einstein's theory does not have any propagating degree of freedom
in three dimensions. In other dimensions, on the other hand, (\ref{denk3})
does have nontrivial solutions. In what follows, since we work explicitly
in three dimensions, we shall gauge away this Einsteinian solution
and not write it. In (\ref{eq:Gen_soln}), each $V_{n}$ satisfies
\begin{equation}
\left(\mathcal{O}+m_{n}^{2}\right)\, V_{n}=0.\label{denk4}
\end{equation}
In the case that two or more coalescing $m_{n}^{2}$'s , the structure
of the solution changes dramatically; for example, the asymptotic
behavior is no longer that of AdS. Let $r$ be the number (multiplicity)
of $m_{n}^{2}$'s that are equal to say $m_{r}^{2}$, then the corresponding
$V_{r}$ satisfies an nonfactorizable higher derivative equation;

\begin{equation}
\left(\mathcal{O}+m_{r}^{2}\right)^{r}\, V_{r}=0.\label{denk5}
\end{equation}
The most general solution now becomes 
\begin{equation}
V=V_{r}+\sum_{n=1}^{N-r}\, V_{n},
\end{equation}
where $V_{r}$ contains $\log^{p}$ terms with $p=1,2,\dots,r-1$.
Such theories are called critical ($r$-critical). Note that $m_{r}^{2}$
may also be zero. Then, the most general solution is in the form
\begin{equation}
V=V_{r0}+\sum_{n=1}^{N-r}\, V_{n},
\end{equation}
where $V_{r0}$ is the solution of $\mathcal{O}^{r+1}\, V_{r0}=0$
and involves $\log^{p}$ terms with $p=1,2,\dots,r$. If all the mass
parameters are all equal to zero, then the criticality reaches its
maximum value of $N+1$. Furthermore, the relation between the maximum
criticality and the derivative order of any gravity theory is worth
mentioning: $\frac{\mbox{derivative order}}{\mbox{maximum criticality}}=2$.
For the case of maximum criticality, the field equations take the
form

\begin{equation}
\mathcal{O}^{N+1}\, V=0.
\end{equation}
As noted above, for the pp-wave metrics, the above discussions are
also valid but in the limit the AdS radius goes to infinity, $\ell\rightarrow\infty$. 

The layout of the paper is as follows: In Sec.~\ref{sec:Sixth-Order-Theory},
we define the most general sixth order theory in three dimensions
and give its field equations for KSK metrics from which the masses
of the spin-2 excitations around the (A)dS background can be obtained.
In Sec.~\ref{sec:AdS-Wave-Solutions}, we give the solutions of the
sixth-order theory and in the ensuing section we extend these solutions
to all higher order derivative theories. In Sec.~\ref{sec:pp-Wave-Solutions},
we also give the pp-wave solutions of sixth-order theories and beyond.

\section{Sixth Order Theory In Three Dimensions\label{sec:Sixth-Order-Theory}}

To give a nontrivial explicit example in full detail, let us consider
the action 
\begin{equation}
I=\frac{1}{\kappa^{2}}\,\int\, d^{3}x\sqrt{-g}\,\Biggl(F\left(R_{\nu}^{\mu}\right)+{\cal L}_{R\square R}\Biggr),
\end{equation}
where at this stage, $F\left(R_{\nu}^{\mu}\right)$ is an arbitrary
differentiable function of the Ricci tensor but not its derivatives
and the second piece in the action constitutes of the two possible
second derivative terms (up to boundary terms): 
\begin{equation}
{\cal L}_{R\square R}=b_{1}\,\nabla_{\mu}\, R\,\nabla^{\mu}\, R+b_{2}\,\nabla_{\rho}\, R_{\alpha\beta}\,\nabla^{\rho}\, R^{\alpha\beta}.
\end{equation}
In \cite{Paulos} (see also \cite{fRicci}), it was shown that the
$F\left(R_{\nu}^{\mu}\right)$ function can be represented more compactly
as $F\left(R_{\nu}^{\mu}\right)=F\left(R,S_{\nu}^{\mu}S_{\mu}^{\nu},S_{\rho}^{\mu}S_{\mu}^{\nu}S_{\nu}^{\rho}\right)$
after the use of Schouten identities to represent higher curvature
scalars in terms of these three curvature scalars, so that the most
general six-derivative theory takes the form%
\footnote{Note that one does not have to use this procedure. A more direct way
would be to work with $R_{\nu}^{\mu}$ is only independent variable,
but for our purposes the laid out method is better since $S_{\mu\nu}\sim\lambda_{\mu}\lambda_{\nu}$.%
} 
\begin{equation}
I=\frac{1}{\kappa^{2}}\,\int\, d^{3}x\sqrt{-g}\,\Biggl(F\left(R,A,B\right)+{\cal L}_{R\square R}\Biggr),\label{eq:Higher_curv_act_in_RAB}
\end{equation}
where we have defined 
\begin{equation}
A\equiv S_{\nu}^{\mu}S_{\mu}^{\nu},\qquad B\equiv S_{\rho}^{\mu}S_{\mu}^{\nu}S_{\nu}^{\rho}.
\end{equation}
Let us write the field equations coming from the variation of (\ref{eq:Higher_curv_act_in_RAB})
in two parts: 
\begin{equation}
E_{\mu\nu}+H_{\mu\nu}=0,\label{eq:Higher_curv_eom}
\end{equation}
where $E_{\mu\nu}$ comes from the $F\left(R,A,B\right)$ part as
\cite{fRicci}

\begin{align}
E_{\mu\nu}= & -\frac{1}{2}g_{\mu\nu}F+2F_{A}S_{\mu}^{\rho}S_{\rho\nu}+3F_{B}S_{\mu}^{\rho}S_{\rho\sigma}S_{\nu}^{\sigma}+\left(\square+\frac{2}{3}R\right)\left(F_{A}S_{\mu\nu}+\frac{3}{2}F_{B}S_{\mu}^{\rho}S_{\rho\nu}\right)\nonumber \\
 & +\left(g_{\mu\nu}\square-\nabla_{\mu}\nabla_{\nu}+S_{\mu\nu}+\frac{1}{3}g_{\mu\nu}R\right)\left(F_{R}-F_{B}S_{\sigma}^{\rho}S_{\rho}^{\sigma}\right)\label{eq:E_mn}\\
 & -2\nabla_{\alpha}\nabla_{(\mu}\left(S_{\nu)}^{\alpha}F_{A}+\frac{3}{2}S_{\nu)}^{\rho}S_{\rho}^{\alpha}F_{B}\right)+g_{\mu\nu}\nabla_{\alpha}\nabla_{\beta}\left(F_{A}S^{\alpha\beta}+\frac{3}{2}F_{B}S^{\alpha\rho}S_{\rho}^{\beta}\right).\nonumber 
\end{align}
Here, the derivatives of the $F$ function are represented as $F_{R}\equiv\frac{\partial F}{\partial R}$,
$F_{A}\equiv\frac{\partial F}{\partial A}$, and $F_{B}\equiv\frac{\partial F}{\partial B}$.
The second part of the field equations, that is $H_{\mu\nu}$, comes
from the variation of ${\cal L}_{R\square R}$ and is given as 
\begin{eqnarray}
H_{\mu\nu} & = & b_{1}\,\left(\nabla_{\mu}\, R\,\nabla_{\nu}\, R-2R_{\mu\nu}\square R-2\left(g_{\mu\nu}\,\square^{2}-\nabla_{\mu}\,\nabla_{\nu}\square\right)R-\frac{1}{2}\, g_{\mu\nu}\,\nabla_{\alpha}\, R\,\nabla^{\alpha}\, R\right)\nonumber \\
 &  & +b_{2}\,\Biggl(\nabla_{\mu}\, R_{\alpha\beta}\,\nabla_{\nu}\, R^{\alpha\beta}-\square^{2}R_{\mu\nu}-g_{\mu\nu}\,\nabla_{\rho}\,\nabla_{\sigma}\square R^{\rho\sigma}+2\nabla^{\rho}\,\nabla_{(\mu}\square R_{\nu)\rho}\nonumber \\
 &  & \phantom{+b_{2}\,\Biggl(}+2\nabla^{\rho}\, R_{\rho\sigma}\,\nabla_{(\mu}\, R_{\nu)}^{\sigma}+2R_{\rho\sigma}\nabla^{\rho}\,\nabla_{(\mu}R_{\nu)}^{\sigma}-2R_{\sigma(\mu}\,\square R_{\nu)}^{\sigma}\nonumber \\
 &  & \phantom{+b_{2}\,\Biggl(}-2\nabla_{\rho}\, R_{\sigma(\mu}\nabla_{\nu)}\, R^{\rho\sigma}-2R_{\sigma(\mu}\nabla^{\rho}\,\nabla_{\nu)}R_{\rho}^{\sigma}-\frac{1}{2}\, g_{\mu\nu}\,\nabla_{\rho}\, R_{\alpha\beta}\,\nabla^{\rho}\, R^{\alpha\beta}\Biggr).\label{eq:Hmn}
\end{eqnarray}

For the metric (\ref{ks}), with the properties listed in (\ref{eq:null}--\ref{eq:Kundt}),
we noted that $S_{\mu\nu}$ is in the form (\ref{tracefree}) and
furthermore, the following identities can be computed from the listed
properties of the metric:
\begin{eqnarray}
 &  & \nabla^{\alpha}\,\square R_{\alpha\mu}=0,~~~\nabla^{\alpha}\, R_{\alpha\mu}=0\\
 &  & \nabla_{\mu}\, R_{\alpha\beta}\,\nabla_{\nu}\, R^{\alpha\beta}=0,\\
 &  & \nabla_{\rho}\, R_{\sigma\mu}\,\nabla_{\nu}\, R^{\rho\sigma}=0,\\
 &  & \nabla^{\rho}\,\nabla_{\mu}\, R_{\rho\nu}=-\frac{3}{\ell^{2}}S_{\mu\nu},\\
 &  & \nabla^{\rho}\,\nabla_{\mu}\square R_{\nu\rho}=-\frac{3}{\ell^{2}}\square S_{\mu\nu},\\
 &  & R^{\rho}\,_{\mu}\,\nabla^{\sigma}\,\nabla_{\nu}\, R_{\rho\sigma}=\frac{6}{\ell^{4}}S_{\mu\nu},\\
 &  & R^{\rho\sigma}\,\,\nabla_{\sigma}\,\nabla_{\mu}\, R_{\nu\rho}=\frac{6}{\ell^{4}}S_{\mu\nu},\\
 &  & R^{\rho}\,_{\mu}\,\square R_{\nu\rho}=-\frac{2}{\ell^{2}}\square S_{\mu\nu}.
\end{eqnarray}
In deriving these identities, we have used the representation of the
three-dimensional Riemann tensor in terms of the Ricci tensor and
the scalar curvature, and also the identity $\lambda^{\rho}\nabla_{\nu}\, S_{\rho\sigma}=0$
which is valid for the KSK class of metrics to which our gravity waves
in AdS belong.

With these identities $H_{\mu\nu}$ reduces to the following from
\begin{equation}
H_{\mu\nu}=-b_{2}\,\Biggl(\square+\frac{2}{\ell^{2}}\Biggr)\square S_{\mu\nu}.
\end{equation}
The metric discussed above represents constant curvature, Type-N spacetimes
as $S_{\mu\nu}$ has the form $S_{\mu\nu}=\rho\lambda_{\mu}\lambda_{\nu}$.
Then, the field equations for these spacetimes (\ref{eq:E_mn}) becomes
\begin{equation}
\left(\frac{1}{3}RF_{R}-\frac{1}{2}F\right)g_{\mu\nu}+\left[-b_{2}\,\square^{2}+\left(F_{A}-\frac{2b_{2}}{\ell^{2}}\right)\square-\frac{1}{3}RF_{A}+F_{R}\right]S_{\mu\nu}=0,\label{eq:EoM_for_KSK}
\end{equation}
where for the $E_{\mu\nu}$ part, results of \cite{fRicci} was used.
The trace of (\ref{eq:EoM_for_KSK}) yields 
\begin{equation}
\frac{1}{3}RF_{R}-\frac{1}{2}F=0,
\end{equation}
which determines the cosmological constant or the AdS radius $\ell$.
The traceless part of (\ref{eq:EoM_for_KSK}) becomes the nonlinear
equation 
\begin{equation}
\left[-b_{2}\,\square^{2}+\left(F_{A}-\frac{2b_{2}}{\ell^{2}}\right)\square-\frac{1}{3}RF_{A}+F_{R}\right]S_{\mu\nu}=0.\label{eq:Unfactorized_EoM}
\end{equation}
which can be rewritten as a product of two operators in general; 
\begin{equation}
\left(\square+\frac{2}{\ell^{2}}-m_{-}^{2}\right)\left(\square+\frac{2}{\ell^{2}}-m_{+}^{2}\right)S_{\mu\nu}=0,\label{eq:Trless_EoM}
\end{equation}
where the mass-squared parameters follow from (\ref{eq:Unfactorized_EoM})
as 
\begin{equation}
m_{\pm}^{2}=\frac{1}{\ell^{2}}+\frac{F_{A}}{2b_{2}}\mp\sqrt{\left(\frac{1}{\ell^{2}}+\frac{F_{A}}{2b_{2}}\right)^{2}+\frac{1}{b_{2}}F_{R}}.\label{eq:m+-2}
\end{equation}
This formula represents the masses of the two spin-2 excitations for
the most general sixth order gravity theory. Once the explicit form
of $F$ is given, one can calculate the masses of these modes. For
example, for the choice of the most general quadratic curvature gravity
in three dimensions, $F\left(R_{\nu}^{\mu}\right)$ has the form 
\begin{equation}
F\left(R_{\nu}^{\mu}\right)=\sigma R-2\lambda_{0}+\alpha R^{2}+\beta R_{\alpha\beta}\, R^{\alpha\beta}=\sigma R-2\lambda_{0}+\left(\alpha+\frac{\beta}{3}\right)R^{2}+\beta S_{\alpha\beta}\, S^{\alpha\beta},
\end{equation}
yielding 
\begin{equation}
F_{R}=\sigma-\frac{12}{\ell^{2}}\left(\alpha+\frac{\beta}{3}\right),\qquad F_{A}=\beta,
\end{equation}
and the square of the mass reads 
\begin{equation}
m_{\pm}^{2}=\frac{\beta}{2b_{2}}+\frac{1}{\ell^{2}}\mp\frac{1}{2b_{2}}\sqrt{\beta^{2}+\frac{4b_{2}^{2}}{\ell^{4}}+4b_{2}\sigma-\frac{12b_{2}}{\ell^{2}}\left(\beta+4\alpha\right)}.\label{eq:m+-2_quad_and_sixth_order}
\end{equation}

For a six-derivative theory, the mass-squared terms, $m_{\pm}^{2}$,
can be arranged to be zero given that $\frac{1}{\ell^{2}}+\frac{F_{A}}{2b_{2}}=0$
and $F_{R}=0$. In this limit, the field equations of the so called
tricritical theories reduce to the form

\begin{equation}
\mathcal{O}^{3}V=0,
\end{equation}
and, hence, have the same logarithmic solutions that we discuss in
the next section.

\section{${\rm AdS}$-Wave Solutions\label{sec:AdS-Wave-Solutions}}

Let us now discuss the exact solutions of (\ref{eq:Trless_EoM}) which
fall into several distinct classes depending on the values of $m_{\pm}^{2}$.
In the generic case, $m_{+}^{2}\ne m_{-}^{2}$. As a second case,
$m_{+}^{2}=m_{-}^{2}\ne0$. In the third case, one of them could be
zero. In the last case, $m_{+}^{2}=m_{-}^{2}=0$.

\paragraph{\uline{Case 1-- $m_{+}^{2}\protect\ne m_{-}^{2}$}:}

For this case, (\ref{eq:Trless_EoM}) reduces to 
\begin{equation}
\left(\mathcal{O}+m_{-}^{2}\right)\left(\mathcal{O}+m_{+}^{2}\right)\mathcal{O}V=0,\label{eq:EoM_in_V}
\end{equation}
whose solutions can be obtained from the solutions of the lower derivative
equations 
\begin{align}
\mathcal{O}V_{E} & =0,\\
\left(\mathcal{O}+m_{+}^{2}\right)V_{+} & =0,\\
\left(\mathcal{O}+m_{-}^{2}\right)V_{-} & =0,
\end{align}
as $V=V_{E}+V_{+}+V_{-}$. Here, $V_{E}$ refers to the solution of
$S_{\mu\nu}=0$. Let us note that $\left(\mathcal{O}+m^{2}\right)\mathcal{O}V=0$
is the traceless part of the field equation for these metrics of the
quadratic curvature gravity, and hence, in some sense for these metrics
the field equations of the sixth order theory reduce to two copies
of the quadratic theory.

With the specific choices of $\lambda^{\mu}$ vector, one can get
the AdS-plane and AdS-spherical wave solutions. The AdS-plane wave
metric can be given as
\begin{equation}
{\rm d}s^{2}=\frac{\ell^{2}}{z^{2}}\left(2{\rm d}u{\rm d}v+{\rm d}z^{2}\right)+2V\left(u,z\right){\rm d}u^{2},\label{eq:AdS-plane_metric}
\end{equation}
where the null coordinates are defined as $u=\frac{1}{\sqrt{2}}\left(x+t\right)$
and $v=\frac{1}{\sqrt{2}}\left(x-t\right)$. Then, the relevant differential
equation becomes \cite{gur-sis-tek}
\begin{equation}
\left(\frac{z^{2}}{\ell^{2}}\frac{\partial^{2}}{\partial z^{2}}+\frac{3z}{\ell^{2}}\frac{\partial}{\partial z}-m_{\pm}^{2}\right)V\left(u,z\right)=0,
\end{equation}
whose solution is
\begin{equation}
V_{\pm}\left(u,z\right)=\frac{1}{z}\left(c_{1}z^{p_{\pm}}+c_{2}z^{-p_{\pm}}\right),\label{eq:AdS-plane_wave_soln}
\end{equation}
where $p_{\pm}\equiv\sqrt{1+m_{\pm}^{2}\ell^{2}}$ and $c_{1,2}$
are functions of $u$. This solution was given in the case of NMG
in \cite{Hassaine}. In $p\rightarrow1$ limit, one obtains the Einsteinian
solution
\begin{equation}
V_{E}\left(u,z\right)=\frac{1}{z^{2}}\left(c_{1}z^{2}+c_{2}\right).\label{eq:VE_AdS-plane}
\end{equation}
From this form, it is easy to see that with this $V_{E}$, (\ref{eq:AdS-plane_metric})
is the AdS space.

The metric in the coordinates used in \cite{gurses1} reads 
\begin{equation}
{\rm d}s^{2}=\frac{\ell^{2}}{\cos^{2}\theta}\left(\frac{4{\rm d}u{\rm d}v}{\left(u+v\right)^{2}}+{\rm d}\theta^{2}\right)+2V\left(u,\theta\right){\rm d}u^{2},\label{eq:AdS-spherical_metric}
\end{equation}
which is called the AdS-spherical wave as the null coordinates are
defined as $u=\frac{1}{\sqrt{2}}\left(r+t\right)$ and $v=\frac{1}{\sqrt{2}}\left(r-t\right)$,
so the AdS part is conformal to the flat space in spherical coordinates.
Then, the relevant differential equation reduces to 
\begin{equation}
\left[\cos^{2}\theta\frac{\partial^{2}}{\partial\theta^{2}}-3\sin\theta\cos\theta\frac{\partial}{\partial\theta}-\left(2\cos^{2}\theta+m_{\pm}^{2}\ell^{2}\right)\right]V_{\pm}\left(u,\theta\right)=0,
\end{equation}
whose solution is 
\begin{equation}
V_{\pm}\left(u,\theta\right)=\frac{1}{\cos\theta}\left[c_{1}\left(\frac{\cos\theta}{1+\sin\theta}\right)^{p_{\pm}}+c_{2}\left(\frac{\cos\theta}{1+\sin\theta}\right)^{-p_{\pm}}\right],\label{eq:V+-_soln}
\end{equation}
where $c_{1,2}$ are arbitrary functions of the null coordinate $u$.
Again, this solution was given in the case of NMG in different coordinates
in \cite{Gurses-Killing,Ahmedov-Aliev-PRL,Ahmedov-Aliev-PLB}. The
$p=1$ case yields the Einsteinian solution 
\begin{equation}
V_{E}\left(u,\theta\right)=\frac{1}{\cos^{2}\theta}\left(c_{+}+c_{-}\sin\theta\right),\label{eq:VE_AdS-spherical}
\end{equation}
where $c_{\pm}=c_{2}\pm c_{1}$ which can again be gauged away. Therefore,
the general AdS-spherical wave solution to the most general sixth
order gravity is 
\begin{equation}
V\left(u,\theta\right)=V_{+}\left(u,\theta\right)+V_{-}\left(u,\theta\right),
\end{equation}
where $V_{+}$ and $V_{-}$ are given in (\ref{eq:V+-_soln}) and
the mass parameters $m_{+}^{2}$ and $m_{-}^{2}$ are given in (\ref{eq:m+-2}).

\paragraph{\uline{Case 2-- $m_{+}^{2}=m_{-}^{2}\protect\ne0$}:}

The AdS-plane wave solution of the fourth order massive operator part
is
\begin{align}
V_{m}=\frac{1}{z} & \left[c_{1}z^{p}+c_{2}z^{-p}+\ln\left(\frac{z}{\ell}\right)\left(c_{3}z^{p}+c_{4}z^{-p}\right)\right],
\end{align}
 while the AdS-spherical wave solution is 
\begin{align}
V_{m}=\frac{1}{\cos\theta} & \Biggl(c_{1}\left(\frac{\cos\theta}{1+\sin\theta}\right)^{p}+c_{2}\left(\frac{\cos\theta}{1+\sin\theta}\right)^{-p}\nonumber \\
 & +\ln\left(\frac{\cos\theta}{1+\sin\theta}\right)\left[c_{3}\left(\frac{\cos\theta}{1+\sin\theta}\right)^{p}+c_{4}\left(\frac{\cos\theta}{1+\sin\theta}\right)^{-p}\right]\Biggr).
\end{align}
The log-terms appear because of the genuinely fourth order nature
of the equation.

\paragraph{\uline{Case 3--one of the masses is zero}:}

The solution is
\begin{equation}
V=V_{m}+V_{{\rm log}},
\end{equation}
where, for the AdS-plane wave, one finds
\begin{equation}
V_{{\rm log}}\left(u,z\right)=\frac{1}{z^{2}}\ln\left(\frac{z}{\ell}\right)\left(c_{3}z^{2}+c_{4}\right),
\end{equation}
also appeared in NMG case \cite{Hassaine}, and for the AdS-spherical
wave, one finds 
\begin{equation}
V_{{\rm log}}\left(u,\theta\right)=\frac{1}{\cos^{2}\theta}\ln\left(\frac{\cos\theta}{1+\sin\theta}\right)\left(c_{3}+c_{4}\sin\theta\right).
\end{equation}

\paragraph{\uline{Case 4-- $m_{+}^{2}=m_{-}^{2}=0$}:}

In this case, the theory is called tricritical \cite{berg1}. The
AdS-plane wave solution is
\begin{equation}
V_{{\rm log}}\left(u,z\right)=\frac{1}{z^{2}}\ln\left(\frac{z}{\ell}\right)\left[c_{3}z^{2}+c_{4}+\ln\left(\frac{z}{\ell}\right)\left(c_{5}z^{2}+c_{6}\right)\right],
\end{equation}
which was partially covered in \cite{Setare}, while the AdS-spherical
wave solution is

\begin{equation}
V_{{\rm log}}\left(u,\theta\right)=\frac{1}{\cos^{2}\theta}\ln\left(\frac{\cos\theta}{1+\sin\theta}\right)\left[c_{3}+c_{4}\sin\theta+\ln\left(\frac{\cos\theta}{1+\sin\theta}\right)\left(c_{5}+c_{6}\sin\theta\right)\right].
\end{equation}

\section{Extension to Any Higher Derivative Order\label{sec:Extension_to_higher}}

\noindent As noted above, for the AdS-wave metrics, the traceless
part of the field equations of any $\left(2N+2\right)$-derivative
theory in three dimensions reduce to the following product

\begin{equation}
\left(\mathcal{O}+m_{1}^{2}\right)\left(\mathcal{O}+m_{2}^{2}\right)\cdots\left(\mathcal{O}+m_{N}^{2}\right)\mathcal{O}V=0,\label{eq:EoM_of_higher_order}
\end{equation}
where $m_{i}$ are the masses of the spin-2 excitations which can
be found in a rather tedious procedure in terms of the parameters
of the theory once the Lagrangian of the theory is given. In Sec.~\ref{sec:Sixth-Order-Theory},
we gave an explicit example for the sixth order gravity. Solutions
of (\ref{eq:EoM_of_higher_order}) depend on whether the masses are
equal or not.

\paragraph{\uline{Case 1--All the masses are distinct}:}

For this case, the most general solution is the sum of the solutions
of each massive operator part as 
\begin{equation}
V=\sum_{i=1}^{N}\, V_{i},
\end{equation}
where the solutions $V_{i}$ are given in (\ref{eq:AdS-plane_wave_soln})
for the AdS-plane wave and in (\ref{eq:AdS-spherical_metric}) for
the AdS-spherical wave. Here, we again dropped the Einsteinian part.

\paragraph{\uline{Case 2--Some masses are equal but not zero}:}

For the case where $r$ number of masses are equal to $m$, the general
solution takes the form
\begin{equation}
V=\sum_{i=0}^{r-1}V_{m}\left(c_{i},c_{i+1}\right)\left(\ln f\right)^{i}+\sum_{i=1}^{N-r}V_{i},
\end{equation}
where $f=\frac{z}{\ell}$ and $V_{m}\left(c_{i},c_{i+1}\right)$ is
given in (\ref{eq:AdS-plane_wave_soln}) for the AdS-plane wave, and
$f=\frac{\cos\theta}{1+\sin\theta}$ and $V_{m}\left(c_{i},c_{i+1}\right)$
is given in (\ref{eq:AdS-plane_wave_soln}) for the AdS-spherical
wave.

\paragraph{\uline{Case 3--some or all of the masses are zero}:}

If $r$ number of masses are zero, then the general solution is 
\begin{equation}
V=\sum_{i=1}^{r}V_{E}\left(c_{i},c_{i+1}\right)\left(\ln f\right)^{i}+\sum_{i=1}^{N-r}V_{i},
\end{equation}
where $f=\frac{z}{\ell}$ and $V_{E}\left(c_{i},c_{i+1}\right)$ is
given in (\ref{eq:VE_AdS-plane}) for the AdS-plane wave, and $f=\frac{\cos\theta}{1+\sin\theta}$
and $V_{E}\left(c_{i},c_{i+1}\right)$ is given in (\ref{eq:VE_AdS-spherical})
for the AdS-spherical wave. When all of the masses are zero, that
is the maximal criticality case, then the general solution becomes
\begin{equation}
V=\sum_{i=1}^{N}V_{E}\left(c_{i},c_{i+1}\right)\left(\ln f\right)^{i}.
\end{equation}

\section{${\rm pp}$-Wave Solutions\label{sec:pp-Wave-Solutions}}

Finally, let us discuss the pp-wave solutions which read in the Kerr-Schild
form as

\begin{equation}
g_{\mu\nu}=\eta_{\mu\nu}+2V\lambda_{\mu}\lambda_{\nu},\label{eq:pp_KS}
\end{equation}
where $\eta_{\mu\nu}$ is the flat Minkowski metric. The function
$V$ satisfies the property $\lambda^{\mu}\partial_{\mu}V=0$. The
vector $\lambda_{\mu}$ is null $\lambda_{\mu}\lambda^{\mu}=0$ and
satisfies $\nabla_{\mu}\lambda_{\nu}=0$.%
\footnote{One may consider the possibility of extending the condition $\nabla_{\mu}\lambda_{\nu}=0$
to the more general condition $\nabla_{\mu}\lambda_{\nu}=\frac{1}{2}\left(\lambda_{\mu}\xi_{\nu}+\xi_{\mu}\lambda_{\nu}\right)$.%
}

It is well-known that for pp-wave spacetimes, the Ricci tensor takes
the form $R_{\mu\nu}=-\lambda_{\mu}\lambda_{\nu}\partial^{2}V$ where
$\partial^{2}$ is the flat Laplacian. As discussed in \cite{gur-sis-tek},
the field equations of the $\left(2N+2\right)$-derivative gravity
theory for the pp-wave metrics reduce to the form
\begin{equation}
\sum_{n=0}^{N}\, a_{n}\,\square^{n}\, R_{\mu\nu}=-\lambda_{\mu}\lambda_{\nu}\sum_{n=0}^{N}\, a_{n}\,\square^{n}\partial^{2}V=0,
\end{equation}
where $\square$ is the Laplacian of the full metric and $a_{n}$
are constants depending on the parameters of the theory. Here, the
first equality follows from $\nabla_{\mu}\lambda_{\nu}=0$. For the
pp-wave spacetimes, a scalar $\phi$ satisfying $\lambda^{\mu}\nabla_{\mu}\phi=0$
also satisfies $\square\phi=\partial^{2}\phi$ and in turn $\lambda^{\mu}\nabla_{\mu}\square\phi=0$
\cite{gur-sis-tek}. Using $\nabla_{\mu}\lambda_{\nu}=0$, together
with these results, it can be shown that $\square^{n}\partial^{2}V=\left(\partial^{2}\right)^{n+1}V$,
so the field equations become 
\begin{equation}
\sum_{n=0}^{N}\, a_{n}\,\left(\partial^{2}\right)^{n}\,\partial^{2}V=0.
\end{equation}
Furthermore, this equation can also be factorized as in the case of
the AdS-wave metrics:
\begin{equation}
\prod_{n=1}^{N}\,\left(\partial^{2}-m_{n,{\rm flat}}^{2}\right)\partial^{2}V=0,\label{eq:EoM_for_pp-wave}
\end{equation}
where $m_{n,{\rm flat}}^{2}$'s are the mass-squared terms for the
massive spin-2 excitations around the flat spacetime. Note that $m_{n,{\rm flat}}^{2}$'s
are related to the $m_{n}^{2}$'s in the limit $\underset{\ell\rightarrow\infty}{{\rm lim}}m_{n}^{2}=m_{n,{\rm flat}}^{2}$.
If one assumes that the all $m_{n,{\rm flat}}^{2}$'s are distinct
, then the most general solution of (\ref{eq:EoM_for_pp-wave}) is
again in the form 
\begin{equation}
V=V_{E}+\sum_{n=1}^{N}\, V_{n},\label{eq:Gen_soln_pp-wave}
\end{equation}
where $V_{E}$ is the Einsteinian solution solving $\partial^{2}V_{E}=0$
and each $V_{n}$ is the massive solution solving $\left(\partial^{2}-m_{n,{\rm flat}}^{2}\right)V_{n}=0$.
For the case of some $m_{n,{\rm flat}}^{2}$'s are equal, the pp-wave
solutions also follow the same pattern discussed for the AdS-wave
solutions at the end of Sec.~\ref{sec:Introduction} after just changing
$\mathcal{O}\rightarrow-\partial^{2}$.

Now, let us find the pp-wave solutions of the sixth order gravity
for the four cases discussed above.

\paragraph{\uline{Case 1-- $m_{+}^{2}\protect\ne m_{-}^{2}$}:}

For this case, the field equation has the form 
\begin{equation}
\left(\partial^{2}-m_{-}^{2}\right)\left(\partial^{2}-m_{+}^{2}\right)\partial^{2}V=0,
\end{equation}
which has the solution $V=V_{E}+V_{+}+V_{-}$ where $V_{E}$, $V_{+}$,
and $V_{-}$ satisfy $\partial^{2}V=0$, $\left(\partial^{2}-m_{+}^{2}\right)V=0$,
and $\left(\partial^{2}-m_{-}^{2}\right)V=0$, respectively. 

To find the explicit solutions, let us write the pp-wave metric in
the null coordinates
\begin{equation}
{\rm d}s^{2}=2{\rm d}u{\rm d}v+{\rm d}z^{2}+2V\left(u,z\right){\rm d}u^{2}.
\end{equation}
Then, the relevant differential equation becomes $\left(\partial_{z}^{2}-m_{\pm}^{2}\right)V_{\pm}\left(u,z\right)=0$
with the solution 
\begin{equation}
V_{\pm}\left(u,z\right)=c_{1}e^{m_{\pm}z}+c_{2}e^{-m_{\pm}z}.\label{eq:pp-wave_soln_for_any_m}
\end{equation}
As we discussed, the Einsteinian part can be gauged away, so there
is no need to consider $V_{E}$.

\paragraph{\uline{Case 2-- $m_{+}^{2}=m_{-}^{2}\protect\ne0$}:}

The pp-wave solution for this case becomes
\begin{equation}
V_{m}=c_{1}e^{m\, z}+c_{2}e^{-m\, z}+z\left(c_{3}e^{m\, z}+c_{4}e^{-m\, z}\right).
\end{equation}

\paragraph{\uline{Case 3--one of the masses is zero}:}

The solution is
\begin{equation}
V=V_{m}+V_{0},
\end{equation}
where 
\begin{equation}
V_{0}\left(u,z\right)=c_{3}z^{3}+c_{4}z^{2}.
\end{equation}

\paragraph{\uline{Case 4-- $m_{+}^{2}=m_{-}^{2}=0$}:}

In this case, the pp-wave solution is
\begin{equation}
V\left(u,z\right)=c_{1}z^{5}+c_{2}z^{4}+c_{3}z^{3}+c_{4}z^{2}.
\end{equation}

For the general case of $\left(2N+2\right)$-derivative theory, we
have the similar cases:

\paragraph{\uline{Case 1--All the masses are distinct}:}

The general solution is

\begin{equation}
V=\sum_{i=1}^{N}\, V_{i},
\end{equation}
where the solutions $V_{i}$ are given in (\ref{eq:pp-wave_soln_for_any_m}).

\paragraph{\uline{Case 2--Some masses are equal but not zero}:}

For the case where $r$ number of masses are equal to $m$, the general
solution is
\begin{equation}
V=\sum_{i=0}^{r-1}V_{m}\left(c_{i},c_{i+1}\right)z^{i}+\sum_{i=1}^{N-r}V_{i},
\end{equation}
where $V_{m}\left(c_{i},c_{i+1}\right)$ is given in (\ref{eq:pp-wave_soln_for_any_m}).

\paragraph{\uline{Case 3--some or all of the masses are zero}:}

If $r$ number of masses are zero, then the general solution is 
\begin{equation}
V=\sum_{i=2}^{2r+1}c_{i}z^{i}+\sum_{i=1}^{N-r}V_{i},
\end{equation}
where $V_{E}\left(c_{i},c_{i+1}\right)=c_{i}+c_{i+1}z$. When all
of the masses are zero, that is the maximal criticality case, then
the general solution becomes
\begin{equation}
V=\sum_{i=2}^{2N+1}c_{i}z^{i}.
\end{equation}

Note that all the $c_{i}$'s appearing in the solutions of this section
are arbitrary functions of $u$.

\section{Conclusions}

\noindent In this work, we studied wave-type exact solutions of any
higher derivative gravity theory in three dimensions. These solutions
also solve the linearized, perturbative, equations for the spin-2
sector as noted below (\ref{eq:c_n_mass_relation}). The field equations
of the most general gravity theory are highly complicated and nonlinear
that, a priori, it is hard to expect any exact solution (besides the
maximally symmetric ones) to be found in closed form. But, rather
remarkably, we found three different wave type solutions AdS-plane,
AdS-spherical, and the pp-wave in any higher derivative theory which,
by the way, do not exist in pure Einstein's gravity in three dimensions.
The exact solutions, as well as the perturbative solutions, are parametrized
by the values of the masses of the spin-2 excitations. Among the solutions,
there are some critical cases that arise when some of the masses vanish
or are equal to each other. For these critical cases, the operators
take a nonfactorizable form and logarithmic terms appear in the solutions
changing the asymptotic structures of the spacetime. As a specific
example, we worked out the details of the most general sixth-order
gravity for which we determined the field equations and the masses
of the two spin-2 excitations explicitly. This example also covers
the recently introduced tricritical gravity in three dimensions. It
is an open question whether there could be other wave solutions in
these theories.

Here, we were mainly interested in finding the exact wave solutions
(in flat and AdS spacetimes) as well as the spin-2 spectrum of the
generic theory while keeping in mind that these solutions, being the
closest cousins of the globally AdS spacetime with the same curvature
invariants as the latter, have potential applications in the AdS\textsubscript{3}/CFT\textsubscript{2}
context. For the generic solutions, where there are no logarithmic
terms, Brown-Henneaux (BH) type boundary conditions \cite{Brown}
are applicable; while for the logarithmic solutions, one needs to
relax these boundary conditions as was already noted in other theories
\cite{Grumiller-Boundary,Hassaine}. We have not studied the properties
(c-charges, etc) of the putative CFT\textsubscript{2} theory, but
it is quite possible that certain theories among the generic set we
have studied will turn out to have a unitary CFT away from the special
points. On the other hand, we expect that generically, the specific
theories with the log terms will lead to non-unitary CFTs. 

While we have studied a large class of gravity theories in 2+1 dimensions
in this work, we have left several theories which need to be mentioned:
to the most general action, one can add the parity-violating Chern-Simons
term to obtain a new class of theories which will be extensions of
Topologically Massive Gravity \cite{TMG}. In principle, it is easy
to extend our solutions to this more general parity violating theory.
It would be interesting to see if such extensions and their chiral
limits lead to viable boundary CFT theories. Finally, as was recently
suggested \cite{Bergshoeff-Minimal,Tekin-MMGcharges}, a theory can
be consistently defined without an action based on the metric alone,
but with field equations, this theory is called the minimal massive
gravity (MMG) with a single massive helicity 2 graviton with the property
that the theory is unitary both in the bulk and on the boundary. Extension
to the two spin-2 case was given in \cite{Tekin-MMG}. Exact solutions
of these theories and their chiral limits were given in \cite{Giribet-MinimalLog,Altas}.

\section{Acknowledgment}

M.~G. and B.~T. are supported by the TÜB\.{I}TAK grant 113F155.
T.~C.~S. thanks The Centro de Estudios Científicos (CECs) where
part of this work was carried out under the support of Fondecyt with
grant 3140127.

\end{document}